# Estimation of Magnetization, Susceptibility and Specific heat for the Two-dimensional Ising Model in a Non-zero Magnetic field

G. Nandhini, M.Vinoth Kumar and M.V. Sangaranarayanan*

Department of Chemistry

Indian Institute of Technology-Madras, Chennai-600036

India

Email : sangara@iitm.ac.in

Fax : +91 44 22570545

## Abstract

The partition functions for two-dimensional nearest neighbour Ising model in a non-zero magnetic field have been derived for finite square lattices with the help of graph theoretical procedures, show-bit algorithm, enumeration of configurations and transfer matrix methods. A new recurrence relation valid for all lattice sizes is proposed. A novel method of computing the critical temperature has been demonstrated. The magnetization and susceptibility for infinite lattices in the presence of magnetic field is estimated. The critical exponent $\delta$ pertaining to the magnetization has also been computed.

**INTRODUCTION**

The analysis of Ising models constitutes a central theme in statistical mechanics and has diverse applications [1-2]. Although the one-dimensional nearest neighbour Ising model did not exhibit phase transitions, the exact solution of the two-dimensional Ising model for zero magnetic field given by Onsager [3] yielded an explicit expression for critical temperatures. Lee and Yang [4] provided a complete solution of the two-dimensional Ising model for vanishing magnetic fields. However, subsequent attempts of solving the two-dimensional Ising models when the magnetic field is non-zero has remained elusive till now; nevertheless methodologies employing Bragg-Williams approximation [5], Bethe *ansatz* [6], series expansions [7], renormalization group [8] scaling hypothesis [9], etc. have been investigated extensively. Due to the equivalence with binary alloys and lattice gas description of fluids [10], the results derived in the context of Ising models become applicable *mutatis mutandis* to various topics in solid state and condensed matter physics. Among several applications of the two-dimensional Ising models [11], mention may be made of the following: order-disorder transitions [12], electrochemical interfaces [13], phase separation in self-assembled monolayer films [14], protein folding [15], free energies of surface steps [16] etc. In contrast to the one-dimensional case, the analysis of two-dimensional Ising models employing the transfer matrix method is considered almost impossible since the total number of spin configurations for a square lattice of N sites is $2^N$ and even for N=16, this number is 65,536. As has been pointed out elsewhere [17] "In terms of the number of papers published, Ising model ranks as probably the most celebrated model in mathematical physics."

In this Communication, we derive the canonical partition function of the two-dimensional nearest neighbour Ising models for a non-zero magnetic field pertaining to a square lattice of 16 sites using (i) enumeration of all the $2^{16}$ configurations and the corresponding energies; (ii) 'show-bit' algorithm (decimal to binary conversion); (iii) graph theoretical approach of counting of 'black-white' edges and (iv) transfer matrix formulation. The partition functions are also deduced for N =25; N=36 and N =64 sites. A general recurrence relation among the partition functions valid for all lattice sizes is also proposed. We also demonstrate a new method of obtaining the critical temperature for finite lattice sizes directly from the energetics of the system, without recourse to the expression for the magnetization. The exact values for the specific heat, magnetization and susceptibilities are reported for N=16, N=36 and N=64 sites for non-zero magnetic field and the extrapolated values for infinite lattices are also provided.

**Methodology**

For the two-dimensional nearest neighbour Ising model on a square lattice [18], the Hamiltonian is given by

$$H_T = -J \sum_{<ij>} (\sigma_{i,j}\sigma_{i,j+1} + \sigma_{i,j}\sigma_{i+1,j}) - H \sum \sigma_{i,j} \quad (1)$$

where J is the nearest neighbour interaction energy, H being the external magnetic field. i and j denote the row and column index respectively.

The periodic boundary conditions are assumed. The corresponding canonical partition function is defined as [19]

$$Q(H,T) = \sum_{i=1}^{2^N} e^{-\frac{(H_T)_i}{kT}} \quad (2)$$

where k denotes the Boltzmann constant, T being the absolute temperature.

**Partition function for a square lattice of 16 sites**

**A. Enumeration of configurations**

By denoting $E_i$ as the energy of $i^{th}$ configuration the partition function is re-written as

$$Q(H,T) = \sum_{i=1}^{2^N} e^{-\frac{E_i}{kT}}$$

where $E_i$ refers to the energy of $i^{th}$ configuration and i varies from 1 to $2^N$.

For brevity, we represent $e^{J/kT}$ as $x$ and $e^{H/kT}$ as $y$; J and H are assumed to be positive quantities. For a square lattice of 16 sites, our *tour de force* consists in deducing the partition function by systematically enumerating all the 65,536 configurations; this task is accomplished by employing Visual Basic programming in conjunction with the Excel software. A typical energy state is as follows:

[$\sigma_{1,1} \times (\sigma_{1,2}+\sigma_{2,1}) + \sigma_{1,2} \times (\sigma_{1,3}+\sigma_{2,2}) + \sigma_{1,3} \times (\sigma_{1,4}+\sigma_{2,3}) + \sigma_{1,4} \times (\sigma_{2,4}+\sigma_{1,1}) +$

$\sigma_{2,1} \times (\sigma_{2,2}+\sigma_{3,1}) + \sigma_{2,2} \times (\sigma_{2,3}+\sigma_{3,2}) + \sigma_{2,3} \times (\sigma_{2,4}+\sigma_{3,3}) + \sigma_{2,4} \times (\sigma_{2,1}+\sigma_{3,4}) +$

$\sigma_{3,1} \times (\sigma_{3,2}+\sigma_{4,1}) + \sigma_{3,2} \times (\sigma_{3,3}+\sigma_{4,2}) + \sigma_{3,3} \times (\sigma_{3,4}+\sigma_{4,3}) + \sigma_{3,4} \times (\sigma_{4,4}+\sigma_{3,1}) +$

$\sigma_{4,1} \times (\sigma_{4,2}+\sigma_{1,1}) + \sigma_{4,2} \times (\sigma_{4,3}+\sigma_{1,2}) + \sigma_{4,3} \times (\sigma_{4,4}+\sigma_{1,3}) + \sigma_{4,4} \times (\sigma_{4,1}+\sigma_{1,4})$]   (3)

wherein the periodic boundary conditions are assumed. In view of this enumeration, the analysis gets simplified as a summation of all the 65536 'energy' terms. The sum

$$\sum_{i=1}^{65536} e^{-\frac{E_i}{kT}}$$ yields the canonical partition function for the square lattice of 16 sites as

$Q_{16} = x^{32}(y^{16} + y^{-16}) + 16x^{24}(y^{14} + y^{-14}) + (88x^{16} + 32x^{20})(y^{12} + y^{-12}) + (96x^{16} + 256x^{12} + 208x^{8})(y^{10} + y^{-10}) + (24x^{16} + 256x^{12} + 736x^{8} + 576x^{4} + 228)(y^{8} + y^{-8}) + (192x^{12} + 688x^{8} + 1664x^{4} + 448x^{-4} + 128x^{-8} + 1248)(y^{6} + y^{-6}) + (96x^{12} + 704x^{8} + 1824x^{4} + 1568x^{-4} + 768x^{-8} + 64x^{-12} + 56x^{-16} + 2928)(y^{4} + y^{-4}) + (64x^{12} + 624x^{8} + 1920x^{4} + 3136x^{-4} + 1392x^{-8} + 512x^{-12} + 96x^{-16} + 16x^{-24} + 3680)(y^{2} + y^{-2}) + (2x^{-32} + 64x^{-20} + 120x^{-16} + 576x^{-12} + 2112x^{-8} + 3264x^{-4} + 1600x^{4} + 768x^{8} + 8x^{16} + 4356)$

(4)

where $x = e^{J/kT}$ and $y = e^{H/kT}$.

**(B) 'Show-bit' algorithm**

The earlier method of summation of individual energy states becomes tedious for larger lattice sizes and an alternate method commonly referred to as 'Show-bit' algorithm is employed herein for the square lattices (N=25, 36). In this algorithm, a decimal number is converted to its corresponding binary number where the 0s are replaced by -1s. The energy of each configuration is estimated and progressively summed in order to compute the corresponding partition function for each N. The partition functions for N=16 are entirely in agreement both numerically and analytically with that obtained from (A). In the case of N =25 and N =36, the energy pertaining to each of the $2^{25}$ and $2^{36}$ configurations becomes computationally complex [20]. Hence MPI (Message Passing Interface) coding was employed using C++ programming in a 64-bit Linux cluster. The partition functions for N=25 and N=36 are given below:

Partition function for N = 25

$$Q_{25,y\neq 1} = x^{50}(y^{25} + y^{-25}) + 25x^{42}(y^{23} + y^{-23}) + (50x^{38} + 250x^{34})(y^{21} + y^{-21}) + (150x^{34} + 850x^{30} + 1300x^{26})$$
$$(y^{19} + y^{-19}) + (25x^{34} + 450x^{30} + 2875x^{26} + 5420x^{22} + 3850x^{18})(y^{17} + y^{-17}) + (210x^{30} + 1650x^{26} + 8850x^{22} +$$
$$18675x^{18} + 17000x^{14} + 6745x^{10})(y^{15} + y^{-15}) + (50x^{30} + 1100x^{26} + 6550x^{22} + 26350x^{18} + 53900x^{14} + 53900x^{10} +$$
$$28150x^{6} + 7100x^{2})(y^{13} + y^{-13}) + (650x^{26} + 4800x^{22} + 24350x^{18} + 75450x^{14} + 134600x^{10} + 135150x^{6} +$$
$$75200x^{2} + 26050x^{-2} + 4450x^{-6})(y^{11} + y^{-11}) + (250x^{26} + 3950x^{22} + 20125x^{18} + 79850x^{14} + 196500x^{10} +$$
$$291850x^{6} + 270675x^{2} + 148550x^{-2} + 54475x^{-6} + 13800x^{-10} + 1550x^{-14})(y^{9} + y^{-9}) + (125x^{26} + 2750x^{22} +$$
$$18075x^{18} + 75450x^{14} + 228425x^{10} + 433650x^{6} + 538325x^{2} + 422350x^{-2} + 220750x^{-6} + 76800x^{-10} + 22225x^{-14} +$$
$$3800x^{-18} + 250x^{-22})(y^{7} + y^{-7}) + (10x^{30} + 2300x^{22} + 15350x^{18} + 72250x^{14} + 239470x^{10} + 530800x^{6} + 789500x^{2} +$$
$$776100x^{-2} + 507100x^{-6} + 231620x^{-10} + 77900x^{-14} + 21500x^{-18} + 4450x^{-22} + 400x^{-26} + 10x^{-30})(y^{5} + y^{-5}) + (100x^{26} +$$
$$1400x^{22} + 14850x^{18} + 68250x^{14} + 241900x^{10} + 589550x^{6} + 966500x^{2} + 1084300x^{-2} + 823500x^{-6} + 428450x^{-10} +$$
$$169600x^{-14} + 52650x^{-18} + 13400x^{-22} + 2700x^{-26} + 250x^{-30})(y^{3} + y^{-3}) + (100x^{26} + 1200x^{22} + 14275x^{18} + 65300x^{14} +$$
$$246500x^{10} + 606000x^{6} + 1068600x^{2} + 1249600x^{-2} + 1027975x^{-6} + 572200x^{-10} + 240800x^{-14} + 79200x^{-18} + 23275x^{-22} +$$
$$4300x^{-26} + 975x^{-30})(y^{1} + y^{-1})$$

Partition function for N =36

$Q_{36,y\neq 1}=x^{72}(y^{36}+y^{-36})+36x^{64}(y^{34}+y^{-34})+(558x^{56}+72x^{60})(y^{32}+y^{-32})+(4908x^{48}+2016x^{52}+216x^{56})(y^{30}+y^{-30})+$
$2(27225x^{40}+23688x^{44}+7308x^{48}+648x^{52}+36x^{56})(y^{28}+y^{-28})+(100332x^{32}+153504x^{36}+95544x^{40}+24480x^{44}+2844x^{48}+$
$288x^{52})(y^{26}+y^{-26})+(252792x^{24}+608112x^{28}+648828x^{32}+336024x^{36}+86544x^{40}+13968x^{44}+1452x^{48}+72x^{52})(y^{24}+y^{-24})+$
$2(442980x^{16}+1548144x^{20}+2564496x^{24}+2282256x^{28}+1118268x^{32}+320976x^{36}+62136x^{40}+7632x^{44}+792x^{48})(y^{22}+y^{-22})+$
$2(546516x^{8}+2601864x^{12}+6240528x^{16}+8636976x^{20}+7164882x^{24}+3595176x^{28}+1169172x^{32}+258408x^{36}+41634x^{40}+4968x^{44}+$
$216x^{48})(y^{20}+y^{-20})+2(480916+2931120x^{4}+9645120x^{8}+19324416x^{12}+24871536x^{16}+20542608x^{20}+11037624x^{24}+4035024x^{28}+$
$1040544x^{32}+203232x^{36}+28368x^{40}+2736x^{44}+36x^{48})(y^{18}+y^{-18})+(308574x^{-8}+2233584x^{-4}+9647748+26428248x^{4}+49293576x^{8}+$
$62584632x^{12}+53905896x^{16}+31673808x^{20}+13086414x^{24}+3950064x^{28}+898164x^{32}+155160x^{36}+19764x^{40}+1224x^{44})(y^{16}+y^{-16})+$
$2(150948x^{-16}+1158048x^{-12}+6331896x^{-8}+22527648x^{-4}+57644748+105570288x^{4}+138281976x^{8}+127874304x^{12}+83537208x^{16}+$
$39356208x^{20}+13794192x^{24}+3683376x^{28}+761616x^{32}+118944x^{36}+13464x^{40}+432x^{44})(y^{14}+y^{-14})+(60768x^{-24}+410832x^{-20}+$
$2765664x^{-16}+12133104x^{-12}+40540428x^{-8}+101625120x^{-4}+190654740+265556736x^{4}+269648586x^{8}+198847464x^{12}+107681508x^{16}+$
$43856208x^{20}+13788678x^{24}+3363048x^{28}+642528x^{32}+93384x^{36}+8820x^{40}+72x^{44}+12x^{48})(y^{12}+y^{-12})+(21600x^{-32}+101520x^{-28}+$
$820224x^{-24}+4192704x^{-20}+17364096x^{-16}+56807136x^{-12}+145123416x^{-8}+287356320x^{-4}+435314160+494848800x^{4}+417699720x^{8}+$
$262657728x^{12}+125152164x^{16}+46237392x^{20}+13410360x^{24}+3051504x^{28}+550836x^{32}+74592x^{36}+5184x^{40}+144x^{44})(y^{10}+y^{-10})+$
$2(6696x^{-40}+18144x^{-36}+173304x^{-32}+951552x^{-28}+4592088x^{-24}+18671616x^{-20}+61849620x^{-16}+165822984x^{-12}+354531024x^{-8}+$
$594585792x^{-4}+769393656+753788160x^{4}+557542728x^{8}+313764192x^{12}+137160000x^{16}+47208816x^{20}+12893472x^{24}+2803824x^{28}+$
$475092x^{32}+60840x^{36}+3456x^{40}+144x^{44})(y^{8}+y^{-8})+(1668x^{-48}+2304x^{-44}+28152x^{-40}+153456x^{-36}+757296x^{-32}+3683376x^{-28}+$
$14895720x^{-24}+51576336x^{-20}+148241772x^{-16}+348192192x^{-12}+657496440x^{-8}+979974720x^{-4}+1131880752+997994592x^{4}+$
$673475688x^{8}+351857376x^{12}+144748116x^{16}+47457792x^{20}+12408816x^{24}+2599344x^{28}+423540x^{32}+50160x^{36}+2808x^{40}+$
$144x^{44})(y^{6}+y^{-6})+(306x^{-56}+144x^{-52}+3456x^{-48}+20448x^{-44}+82404x^{-40}+443016x^{-36}+2091492x^{-32}+8660952x^{-28}+32060817x^{-24}+$
$100316088x^{-20}+263218752x^{-16}+566862840x^{-12}+985754664x^{-8}+1356521904x^{-4}+1450906884+1193279472x^{4}+758452707x^{8}+$
$377578800x^{12}+149403996x^{16}+47250864x^{20}+12095820x^{24}+2425752x^{28}+395568x^{32}+42120x^{36}+2700x^{40}+144x^{44})(y^{4}+y^{-4})+$
$2(36x^{-64}+216x^{-56}+2304x^{-52}+6912x^{-48}+33408x^{-44}+182016x^{-40}+806544x^{-36}+3589416x^{-32}+14213952x^{-28}+49325472x^{-24}+$
$146534832x^{-20}+364675500x^{-16}+747933120x^{-12}+1240315488x^{-8}+1628668800x^{-4}+1668054600+1317182688x^{4}+810164736x^{8}+$
$392376672x^{12}+151680312x^{16}+47161728x^{20}+11825928x^{24}+2347344x^{28}+373176x^{32}+38592x^{36}+2664x^{40}+144x^{44})(y^{2}+y^{-2})+$
$(2x^{-72}+144x^{-60}+576x^{-56}+1152x^{-52}+11076x^{-48}+47520x^{-44}+205542x^{-40}+1004256x^{-36}+4302864x^{-32}+16566624x^{-28}+56924808x^{-24}+$
$165466512x^{-20}+405640980x^{-16}+818194560x^{-12}+1336101246x^{-8}+1728000432x^{-4}+1744323016+1359828000x^{4}+827402400x^{8}+$
$397078512x^{12}+152525196x^{16}+47010240x^{20}+11788038x^{24}+2299680x^{28}+373752x^{32}+34704x^{36}+3456x^{40}+12x^{48})$

In view of the computational complexity, the partition function for N=64 was not attempted using the show-bit algorithm.

### (C) Combinatorial approach

The connection of the Ising model problem to the combinatorial graph theory is well known [21]. The essential premise in this context is that the two-dimensional Ising model can be visualized as a bipartite graph G consisting of the set of vertices V and edges E viz. $G = \{V,E\}$. The vertices represent the arrangement of $+1$ ('black') and $-1$ ('white') spins and the edges correspond to the connectivity between the vertices. The black-white edges correspond to the perfect matching of the graph. The number of ways in which a black - white edge can arise, correspond to its weight. This strategy is analogous to the weighted graph colorings problem wherein adjacent vertices have different colours [22].

For brevity, let p denote the number of +1 spins (black vertex) and q represent the black-white edge. Thus, the estimation of the partition function in non-zero magnetic field is now transformed as counting of the black-white edges of a square lattice of N sites wherein adjacent and non-adjacent black-white edges along with their identities need to be included. along with the periodic boundary conditions.

As an illustration for a square lattice of N sites, the number of black sites may be denoted as p and hence that of white sites is N-p. Let q be the number of black-white edges formed. Then, A(p,q) denotes the number of times a configuration occurs with p number of black sites and q number of black-white edges. For example, when there are 2 adjacent black sites, the number of black-white edges will be 6, taking into account, the periodic boundary conditions, although the 2 black sites can occur at any place among the 4 rows and 4 columns(for N =16). The value of p may vary from 0 to N since there can be no black sites, 1 black site, 2 black sites etc. Accordingly, the value of q also changes. The number of black-white edges (q) changes from 0 to 2N and is also dictated by the value of p, since for a given value of p, the black sites can be adjacent or non-adjacent. For example, if p=3 (3 black sites), the following cases arise: (i) all the three can be adjacent, (ii) two of them can be adjacent and the remaining site is present elsewhere within the square and (iii) all the three may be scattered. Table 1 denotes A(p,q) for a square lattice of 16 sites deduced using a program written in C++.

Table 1 A(p,q) values which arise from the counting of 'black-white' edges for a square lattice of 16 sites leading to the partition function

| q \ p | 0 | 1 | 2 | 3 | 4 | 5 | 6 | 7 | 8 | 9 | 10 | 11 | 12 | 13 | 14 | 15 | 16 |
|---|---|---|---|---|---|---|---|---|---|---|---|---|---|---|---|---|---|
| 0 | 1 | | | | | | | | | | | | | | | | 1 |
| 2 | | | | | | | | | | | | | | | | | |
| 4 | | 16 | | | | | | | | | | | | | | 16 | |
| 6 | | | 32 | | | | | | | | | | | | 32 | | |
| 8 | | | 88 | 96 | 24 | | | | 8 | | | | 24 | 96 | 88 | | |
| 10 | | | | 256 | 256 | 192 | 96 | 64 | 0 | 64 | 96 | 192 | 256 | 256 | | | |
| 12 | | | | 208 | 736 | 688 | 704 | 624 | 768 | 624 | 704 | 688 | 736 | 208 | | | |
| 14 | | | | | 576 | 1664 | 1824 | 1920 | 1600 | 1920 | 1824 | 1664 | 576 | | | | |
| 16 | | | | | 228 | 1248 | 2928 | 3680 | 4356 | 3680 | 2928 | 1248 | 228 | | | | |
| 18 | | | | | | 448 | 1568 | 3136 | 3264 | 3136 | 1568 | 448 | | | | | |
| 20 | | | | | | 128 | 768 | 1392 | 2112 | 1392 | 768 | 128 | | | | | |
| 22 | | | | | | | 64 | 512 | 576 | 512 | 64 | | | | | | |
| 24 | | | | | | | 56 | 96 | 120 | 96 | 56 | | | | | | |
| 26 | | | | | | | | 0 | 64 | 0 | | | | | | | |
| 28 | | | | | | | | 16 | 0 | 16 | | | | | | | |
| 30 | | | | | | | | | 0 | | | | | | | | |
| 32 | | | | | | | | | 2 | | | | | | | | |

The number of black-white edges A(p,q) is related to the canonical partition function in two dimensions as [23]

$$Q(x,y) = \sum_{p,q} A(p,q) x^{2(N-q)} y^{N-2p} \qquad (5)$$

where $x=e^{J/kT}$ and $y=e^{H/kT}$ as before. Incorporating A(p,q) and the corresponding p and q values for N =16, the partition function is derived for N =16. This equation is identical with that arising from the other two approaches (A) and (B) mentioned above. The methodology has also been extended to N =25 (Table 2)and N =36 (Appendix).

Table -2 -Counting of the black-white edges for deducing the partition function for a square lattice of 25 sites

| p\q | 0 | 1 | 2 | 3 | 4 | 5 | 6 | 7 | 8 | 9 | 10 | 11 | 12 |
|---|---|---|---|---|---|---|---|---|---|---|---|---|---|
| 0 | 1 | | | | | | | | | | | | |
| 2 | | | | | | | | | | | | | |
| 4 | | 25 | | | | | | | | | | | |
| 6 | | | 50 | | | | | | | | | | |
| 8 | | | 250 | 150 | 25 | | | | | | | | |
| 10 | | | | 850 | 450 | 210 | 50 | | | | 10 | | |
| 12 | | | | 1300 | 2875 | 1650 | 1100 | 650 | 250 | 125 | 0 | 100 | 100 |
| 14 | | | | | 5450 | 8850 | 6550 | 4800 | 3950 | 2750 | 2300 | 1400 | 1200 |
| 16 | | | | | 3850 | 18675 | 26350 | 24350 | 20125 | 18075 | 15350 | 14850 | 14275 |
| 18 | | | | | | 17000 | 53900 | 75450 | 79850 | 75450 | 72250 | 68250 | 65300 |
| 20 | | | | | | 6745 | 53900 | 134600 | 196500 | 228425 | 239470 | 241900 | 246500 |
| 22 | | | | | | | 28150 | 135150 | 291850 | 433650 | 530800 | 589550 | 606000 |
| 24 | | | | | | | 7100 | 75200 | 270675 | 538325 | 789500 | 966500 | 1068600 |
| 26 | | | | | | | | 26050 | 148550 | 422350 | 776100 | 1084300 | 1249600 |
| 28 | | | | | | | | 4450 | 54475 | 220750 | 507100 | 823500 | 1027975 |
| 30 | | | | | | | | | 13800 | 76800 | 231620 | 428450 | 572200 |
| 32 | | | | | | | | | 1550 | 22225 | 77900 | 169600 | 240800 |
| 34 | | | | | | | | | | 3800 | 21500 | 52650 | 79200 |
| 36 | | | | | | | | | | 250 | 4450 | 13400 | 23275 |
| 38 | | | | | | | | | | | 400 | 2700 | 4300 |
| 40 | | | | | | | | | | | 10 | 250 | 975 |
| 42 | | | | | | | | | | | | | |
| 44 | | | | | | | | | | | | | |
| 46 | | | | | | | | | | | | | |
| 48 | | | | | | | | | | | | | |
| 50 | | | | | | | | | | | | | |

| p\q | 13 | 14 | 15 | 16 | 17 | 18 | 19 | 20 | 21 | 22 | 23 | 24 | 25 |
|---|---|---|---|---|---|---|---|---|---|---|---|---|---|
| 0 | | | | | | | | | | | | | 1 |
| 2 | | | | | | | | | | | | | |
| 4 | | | | | | | | | | | | 25 | |
| 6 | | | | | | | | | | | 50 | | |
| 8 | | | | | | | | | 25 | 150 | 250 | | |
| 10 | | | 10 | | | | 50 | 210 | 450 | 850 | | | |
| 12 | 100 | 100 | 0 | 125 | 250 | 650 | 1100 | 1650 | 2875 | 1300 | | | |
| 14 | 1200 | 1400 | 2300 | 2750 | 3950 | 4800 | 6550 | 8850 | 5450 | | | | |
| 16 | 14275 | 14850 | 15350 | 18075 | 20125 | 24350 | 26350 | 18675 | 3850 | | | | |
| 18 | 65300 | 68250 | 72250 | 75450 | 79850 | 75450 | 53900 | 17000 | | | | | |
| 20 | 246500 | 241900 | 239470 | 228425 | 196500 | 134600 | 53900 | | | | | | |
| 22 | 606000 | 589550 | 530800 | 433650 | 291850 | 135150 | 28150 | | | | | | |
| 24 | 1068600 | 966500 | 789500 | 538325 | 270675 | 75200 | 7100 | | | | | | |
| 26 | 1249600 | 1084300 | 776100 | 422350 | 148550 | 26050 | | | | | | | |
| 28 | 1027975 | 823500 | 507100 | 220750 | 54475 | 4450 | | | | | | | |
| 30 | 572200 | 428450 | 231620 | 76800 | 13800 | | | | | | | | |
| 32 | 240800 | 169600 | 77900 | 22225 | 1550 | | | | | | | | |
| 34 | 79200 | 52650 | 21500 | 3800 | | | | | | | | | |
| 36 | 23275 | 13400 | 4450 | 250 | | | | | | | | | |
| 38 | 4300 | 2700 | 400 | | | | | | | | | | |
| 40 | 975 | 250 | 10 | | | | | | | | | | |
| 42 | | | | | | | | | | | | | |
| 44 | | | | | | | | | | | | | |
| 46 | | | | | | | | | | | | | |
| 48 | | | | | | | | | | | | | |
| 50 | | | | | | | | | | | | | |

This enumeration method for counting of vertices and edges has tremendous potentialities and among them, mention may be made of the following: protein folding [24], stability of polycyclic benzenoid hydrocarbons [25], organization of biochemical networks [26], percolation on two dimensional lattices [27] etc.

**(D) Transfer matrix formulation**

For one-dimensional Ising models, the basic transfer matrix turns out to be 2×2, whose eigenvalues can then be employed for all lattice sizes. This simplicity is no longer present for the two-dimensional case. Furthermore, for a square lattice of $n^2$ sites, the transfer matrix [28] is of dimension $2^n \times 2^n$ where n is the number of rows or columns, thus implying that the matrix needs to be constructed separately for each lattice size in general.

It is of interest to point out that the procedure of Kramers and Wanniers [29] for two-dimensional Ising models results in a double-shifted circulant matrix and is valid irrespective of the lattice size albeit at H =0. Inspired by this rigorous methodology, it is tempting to formulate a matrix version for H ≠ 0 too. In the case of one-dimensional Ising models, a Toeplitz matrix does exist [30] for H ≠ 0 and J ≠ 0 which can be obtained from the Discrete Fourier Transformation [31]. However, for two-dimensional Ising models, this computation is tedious even for N =16. Although the methodology for constructing such Toeplitz matrices for N ≥16 is quite tedious, this task when accomplished renders it possible to exploit well-known theorems pertaining to asymptotic limits for eigenvalues of circulant [32] matrices. Thus, the thermodynamic limit of N → ∞ for non-zero magnetic field may possibly be reached from the eigenvalues of the circulant matrices. If the correct matrix for a square lattice of N sites is constructed, the partition function follows from

$$Q(H,T) = \sum_{i=1}^{2^n} (\lambda_i)^n \qquad (6)$$

where n denotes the number of rows or columns.

While there are 65,536 configurations for a square lattice of 4×4 sites, the transfer matrix is indeed a 16×16 matrix consisting of 256 elements. The ferreting out of 256 elements out of the entire set of 65,536 elements in order to construct the transfer matrix appears at first sight to be daunting. However, as has been demonstrated by Kramers and Wanniers [29] there exists an inherent symmetry of the two-dimensional Ising models which can be exploited. Out of $2^N$ configurations, only those configurations which have reflection symmetry are constituents of the matrix. If the condition, site (i) = site (N/2+i) is satisfied in a configuration, the configuration is said to obey the reflection symmetry. For example, in a square lattice of 16 sites, among all the 65536 elements, only 256 elements satisfy this reflection symmetry property and become the constituents of the matrix, i.e., only those configurations, wherein, site(1)=site(9), site (2)= site (10) and site (8)= site (16) etc satisfy the reflection symmetry and constitute the elements of the matrix. As an illustration, the following arrangement denotes the 2314$^{th}$ configuration out of 65536 for N=16.

|  1  |  1  |  1  |  1  | -1  |  1  |  1  | -1  |  1  |  1  |  1  |  1  | -1  |  1  |  1  | -1  |
|-----|-----|-----|-----|-----|-----|-----|-----|-----|-----|-----|-----|-----|-----|-----|-----|
| (1) | (2) | (3) | (4) | (5) | (6) | (7) | (8) | (9) |(10) |(11) |(12) |(13) |(14) |(15) |(16) |

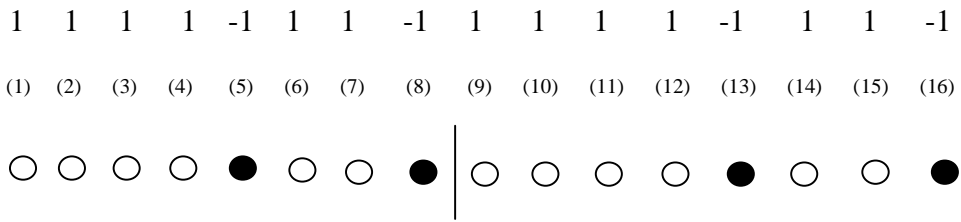

The energy pertaining to the above configuration is $E_{2314}$= -8J -8H.

Hence $e^{-E_{2314}/kT} = e^{8J/kT+8H/kT} = x^8 y^8$. Subsequently, equation (3) is employed to obtain all the coefficients of J and H. After calculating the energies of all the 256 reflective symmetric configurations and thus the corresponding expressions, the first 16 elements are arranged in the first row, the next 16 in the second row and so on, till the last 16 elements constitute the last row, thus leading to a 16×16 matrix. The matrix is such that the upper and lower triangular portions of the main diagonal are symmetric. In view of the matrix being formulated from the 256 elements, the energies x and y need to be scaled as $x_1=x^{1/n}$ and $y_1=y^{1/n}$. It should be emphasized that for a lattice of 16 sites, if all the $2^{16}$ elements are arranged as a $2^8 \times 2^8$ matrix, the 256 diagonal elements are same as the 256 matrix elements which arise from the reflection symmetry method. This is obvious since the partition function is the trace of the matrix corresponding to the Hamiltonian. The matrix pertaining to this lattice is provided in Figure 1. When H =0, i.e y =1, this matrix will correspond to the zero magnetic field case. Thus, it follows that the matrix for a square lattice of 16 sites pertaining to the two-dimensional nearest neighbour Ising model for a non-zero magnetic field along with the algebraic expression for the corresponding partition function is not unattainable! The partition function *equation* deduced from the matrix is identical with that arising from the approaches (A),(B) and (C).When N is an odd number, this reflection property method becomes inapplicable and hence Q could not be obtained for N=25. In the case of N= 64, the $2^{64}$ elements are first arranged into a $2^{32} \times 2^{32}$ matrix, which is further reduced to $2^{16} \times 2^{16}$ using the diagonal elements of the former and then expressed as the $2^{8} \times 2^{8}$ using C++ programming. The expression for the partition function could not be obtained for the large symbolic matrix. However, for various values of H and J, the numerical value of the partition function was obtained easily (Figure 2).But the same procedure cannot be applied to arrive at a matrix for 36 sites. It is well known that the matrix dimension that corresponds to 36 sites is $2^{6} \times 2^{6}$, i.e., the number of matrix elements should be $2^{12}$, i.e., 4096, out of $2^{36}$ elements. Hence, the various combinations obtained from 12 (i.e., 2n) +1s and -1s are reflected 3 times to get 36 +1s and -1s combinations such that site (1) =site (13) =site (25), site (2) = site (14) = site (26) and so on, till, site (12) = site (24) = site (36), resulting in $2^{12}$ total configurations, and the corresponding elements arranged in $2^{6} \times 2^{6}$ matrix.

$$M = \begin{bmatrix}
x_1^{32}y_1^{16} & x_1^{16}y_1^{12} & x_1^{16}y_1^{12} & x_1^{8}y_1^{8} & x_1^{16}y_1^{12} & y_1^{8} & x_1^{8}y_1^{8} & y_1^{4} & x_1^{16}y_1^{12} & x_1^{8}y_1^{8} & y_1^{8} & y_1^{4} & x_1^{8}y_1^{8} & y_1^{4} & y_1^{4} & 1 \\
x_1^{16}y_1^{12} & x_1^{16}y_1^{8} & y_1^{8} & x_1^{8}y_1^{4} & y_1^{8} & y_1^{4} & \frac{y_1^{4}}{x_1^{8}} & 1 & y_1^{8} & x_1^{8}y_1^{4} & \frac{y_1^{4}}{x_1^{16}} & 1 & \frac{y_1^{4}}{x_1^{8}} & 1 & x_1^{-16} & y_1^{-4} \\
x_1^{16}y_1^{12} & y_1^{8} & x_1^{16}y_1^{8} & x_1^{8}y_1^{4} & y_1^{8} & \frac{y_1^{4}}{x_1^{16}} & x_1^{8}y_1^{4} & 1 & y_1^{8} & \frac{y_1^{4}}{x_1^{8}} & y_1^{4} & 1 & \frac{y_1^{4}}{x_1^{8}} & x_1^{-16} & 1 & y_1^{-4} \\
x_1^{8}y_1^{8} & x_1^{8}y_1^{4} & x_1^{8}y_1^{4} & x_1^{16} & \frac{y_1^{4}}{x_1^{8}} & x_1^{-8} & 1 & \frac{x_1^{8}}{y_1^{4}} & \frac{y_1^{4}}{x_1^{8}} & 1 & x_1^{-8} & \frac{x_1^{8}}{y_1^{4}} & x_1^{-16} & \frac{1}{x_1^{8}y_1^{4}} & \frac{1}{x_1^{8}y_1^{4}} & \frac{x_1^{8}}{y_1^{8}} \\
x_1^{16}y_1^{12} & y_1^{8} & y_1^{8} & \frac{y_1^{4}}{x_1^{8}} & x_1^{16}y_1^{8} & y_1^{4} & x_1^{8}y_1^{4} & 1 & y_1^{8} & \frac{y_1^{4}}{x_1^{8}} & \frac{y_1^{4}}{x_1^{16}} & x_1^{-16} & x_1^{8}y_1^{4} & 1 & 1 & y_1^{-4} \\
y_1^{8} & y_1^{4} & \frac{y_1^{4}}{x_1^{16}} & x_1^{-8} & y_1^{4} & 1 & x_1^{-8} & y_1^{-4} & \frac{y_1^{4}}{x_1^{16}} & x_1^{-8} & x_1^{-32} & \frac{1}{x_1^{16}y_1^{4}} & x_1^{-8} & y_1^{-4} & \frac{1}{x_1^{16}y_1^{4}} & y_1^{-8} \\
x_1^{8}y_1^{8} & \frac{y_1^{4}}{x_1^{8}} & x_1^{8}y_1^{4} & 1 & x_1^{8}y_1^{4} & x_1^{-8} & x_1^{16} & \frac{x_1^{8}}{y_1^{4}} & \frac{y^{4}}{x_1^{8}} & x_1^{-16} & x_1^{-8} & \frac{1}{x_1^{8}y_1^{4}} & 1 & \frac{1}{x_1^{8}y_1^{4}} & \frac{x_1^{8}}{y_1^{4}} & \frac{x_1^{8}}{y_1^{8}} \\
y_1^{4} & 1 & 1 & \frac{x_1^{8}}{y_1^{4}} & 1 & y_1^{-4} & \frac{x_1^{8}}{y_1^{4}} & \frac{x_1^{16}}{y_1^{8}} & x_1^{-16} & \frac{1}{x_1^{8}y_1^{4}} & \frac{1}{x_1^{16}y_1^{4}} & y_1^{-8} & \frac{1}{x_1^{8}y_1^{4}} & y_1^{-8} & y_1^{-8} & \frac{x_1^{16}}{y_1^{12}} \\
x_1^{16}y_1^{12} & y_1^{8} & y_1^{8} & \frac{y_1^{4}}{x_1^{8}} & y_1^{8} & \frac{y_1^{4}}{x_1^{16}} & \frac{y_1^{4}}{x_1^{8}} & x_1^{-16} & x_1^{16}y_1^{8} & x_1^{8}y_1^{4} & y_1^{4} & 1 & x_1^{8}y_1^{4} & 1 & 1 & y_1^{-4} \\
x_1^{8}y_1^{8} & x_1^{8}y_1^{4} & \frac{y_1^{4}}{x_1^{8}} & 1 & \frac{y_1^{4}}{x_1^{8}} & x_1^{-8} & x_1^{-16} & \frac{1}{x_1^{8}y_1^{4}} & x_1^{8}y_1^{4} & x_1^{16} & x_1^{-8} & \frac{x_1^{8}}{y_1^{4}} & 1 & \frac{x_1^{8}}{y_1^{4}} & \frac{1}{x_1^{8}y_1^{4}} & \frac{x_1^{8}}{y_1^{8}} \\
y_1^{8} & \frac{y_1^{4}}{x_1^{16}} & y_1^{4} & x_1^{-8} & \frac{y_1^{4}}{x_1^{16}} & x_1^{-32} & x_1^{-8} & \frac{1}{x_1^{16}y_1^{4}} & y^{4} & x_1^{-8} & 1 & y_1^{-4} & x_1^{-8} & \frac{1}{x_1^{16}y_1^{4}} & y_1^{-8} & y_1^{-8} \\
y_1^{4} & 1 & 1 & \frac{x_1^{8}}{y_1^{4}} & x_1^{-16} & \frac{1}{x_1^{16}y_1^{4}} & \frac{1}{x_1^{8}y_1^{4}} & y_1^{-8} & 1 & \frac{x_1^{8}}{y_1^{4}} & y_1^{-4} & \frac{x_1^{16}}{y_1^{8}} & \frac{1}{x_1^{8}y_1^{4}} & y_1^{-8} & y_1^{-8} & \frac{x_1^{16}}{y_1^{12}} \\
x_1^{8}y_1^{8} & \frac{y_1^{4}}{x_1^{8}} & \frac{y_1^{4}}{x_1^{8}} & x_1^{-16} & x_1^{8}y_1^{4} & x_1^{-8} & 1 & \frac{1}{x_1^{8}y_1^{4}} & x_1^{8}y_1^{4} & 1 & x_1^{-8} & \frac{1}{x_1^{8}y_1^{4}} & x_1^{16} & \frac{x_1^{8}}{y_1^{4}} & \frac{x_1^{8}}{y_1^{4}} & \frac{x_1^{8}}{y_1^{8}} \\
y_1^{4} & 1 & x_1^{-16} & \frac{1}{x_1^{8}y_1^{4}} & 1 & y_1^{-4} & \frac{1}{x_1^{8}y_1^{4}} & y_1^{-8} & 1 & \frac{x_1^{8}}{y_1^{4}} & \frac{1}{x_1^{16}y_1^{4}} & y_1^{-8} & \frac{x_1^{8}}{y_1^{4}} & \frac{x_1^{16}}{y_1^{8}} & y_1^{-8} & \frac{x_1^{16}}{y_1^{12}} \\
y_1^{4} & x_1^{-16} & 1 & \frac{1}{x_1^{8}y_1^{4}} & 1 & \frac{1}{x_1^{16}y_1^{4}} & \frac{x_1^{8}}{y_1^{4}} & y_1^{-8} & 1 & \frac{1}{x_1^{8}y_1^{4}} & y_1^{-4} & y_1^{-8} & \frac{x_1^{8}}{y_1^{4}} & y_1^{-8} & \frac{x_1^{16}}{y_1^{8}} & \frac{x_1^{16}}{y_1^{12}} \\
1 & y_1^{-4} & y_1^{-4} & \frac{x_1^{8}}{y_1^{8}} & y_1^{-4} & y_1^{-8} & \frac{x_1^{8}}{y_1^{4}} & \frac{x_1^{16}}{y_1^{12}} & y_1^{-4} & \frac{x_1^{8}}{y_1^{8}} & y_1^{-8} & \frac{x_1^{16}}{y_1^{12}} & \frac{x_1^{8}}{y_1^{8}} & \frac{x_1^{16}}{y_1^{12}} & \frac{x_1^{16}}{y_1^{12}} & \frac{x_1^{32}}{y_1^{16}}
\end{bmatrix}$$

Figure 1 The matrix constructed using the reflective symmetry principle for the square lattice of 16 sites for the two-dimensional Ising model in the presence of the magnetic field.

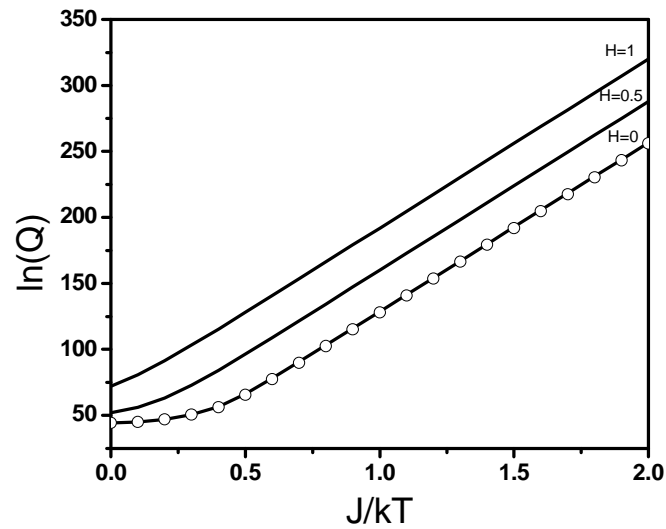

Figure 2 Estimated partition functions for N =64 at various values of H. For H =0, the points denote the values predicted by the Onsager's exact solution while the line arises from the eigenvalues of the 256×256 matrix.

**Results and Discussion**

The foregoing analysis demonstrates that the partition function for a square lattice of 16 sites can be obtained by different methods even in the presence of a magnetic field. However, difficulties emerge for the next square lattice viz N = 25. In this case, the matrix method by reflection symmetry becomes invalid since the number of sites is odd; further, the enumeration procedure described in (A) is tedious. The combinatorial approach for N=25 using A(p,q) values is feasible. The show bit algorithm is convenient for N =16, N =25 and N =36; however, the matrix formulation enables the estimation of the partition function for N =64 but a closed form expression could not be obtained.

**Recurrence Relation**

Thus, the computation of the partition function becomes more complex as the lattice size increases and hence it is preferable to have a valid generalization for large lattices so that the conversion into an integral may become feasible. A careful scrutiny of the partition function estimates pertaining to N =16 and N = 25 reveals a recurrence relation as follows:

$$\ln Q_{(n+2)^2} = \ln Q_{(n+1)^2} + \frac{2n+3}{2n+1}\left(\ln Q_{(n+1)^2} - \ln Q_{n^2}\right) \qquad (7)$$

where n is the number of rows or columns. The partition functions predicted from the above equation for N =36 are almost identical with the values arising directly from the partition function equation for N =36 for all values of H and J. (In the numerical calculations for the partition function, magnetization, specific heat as well as the susceptibility, the magnetic field has been non-dimensionalised with respect to kT). Thus, the partition function for a chosen $(n+2)^2$ value is obtainable from those pertaining to $(n+1)^2$ and $n^2$ sizes. Furthermore, the partition functions for N =16 and N =25 alone are adequate to deduce the partition functions for any other value of N. The free energy for 320*320 lattice for H=0, at the critical temperature Tc (J/ kTc = 0.4407) predicted from the above recurrence relation is 0.929429 while the corresponding value reported by Roland et al [33] is 0.929701. The computed values of lnQ for N = 320*320 from the recurrence relation employing the values for N = 16 and N =25 are provided in Table 3 for H = 0 and H = 0.1.

Table 3 The partition function for a square lattice of 102400 sites arising from the recurrence relation (7) commencing from the partition functions for N=16 and N =25.

| J/kT | lnQ320*320 for H=0 | lnQ320*320 for H=0.1 |
|---|---|---|
| 0.00 | 70978.30 | 71489.40 |
| 0.10 | 72002.10 | 72800.60 |
| 0.20 | 75067.60 | 76522.20 |
| 0.30 | 80288.30 | 83663.70 |
| 0.40 | 89512.60 | 96530.30 |
| 0.50 | 105047.00 | 114117.00 |
| 0.60 | 123909.00 | 133516.00 |
| 0.70 | 143790.00 | 153554.00 |
| 0.80 | 164026.00 | 173847.00 |
| 0.90 | 184401.00 | 194246.00 |
| 1.00 | 204836.00 | 214690.00 |
| 1.10 | 225297.00 | 235154.00 |
| 1.20 | 245768.00 | 255628.00 |
| 1.30 | 266244.00 | 276104.00 |
| 1.40 | 286722.00 | 296583.00 |
| 1.50 | 307201.00 | 317062.00 |
| 1.60 | 327681.00 | 337542.00 |
| 1.70 | 348161.00 | 358022.00 |
| 1.80 | 368641.00 | 378502.00 |
| 1.90 | 389121.00 | 398982.00 |
| 2.00 | 409601.00 | 419462.00 |

**Spontaneous Magnetization and critical temperature**

Critical Temperature of finite lattices from energy considerations

It is well-known that the spontaneous magnetization vanishes for any finite sized lattice [34]. Hence, despite the availability of the partition functions for H ≠ 0 and J ≠ 0 pertaining to N =16, 25, 36, 64 etc, the same is not adequate to deduce the critical temperature from magnetization. In view of this limitation, an alternate procedure for estimating the critical temperature without resorting to the magnetization route is suggested here. Since the critical temperature refers to the temperature at which the spontaneous magnetization changes from +1 (or -1) to 0, it is legitimate to equate the differences in their Helmholtz free energies ($A_N$) for M =1 (all spins +1) and M =0( equal +1 and -1 spins) to zero viz

$A_N \{M=1\} - A_N \{M=0\} = 0$

From the final expression for the partition functions which is a sum over energy states, it is not possible to ferret out the energies with the chosen spin arrangements, even for N =16, implying that the *complete algebraic expressions for Q,* even if available for large lattice size will be *inadequate* unless the individual terms are identified with the configurational states. In the present analysis, the graph theoretical procedure, show-bit algorithm and enumeration of individual energies using EXCEL have the feature wherein the energies corresponding to each arrangement can be deciphered. We illustrate the procedure for deducing Tc from the counting of black-white edges using A(p,q) values of Table 1, pertaining to a square lattice of 16 sites. The spontaneous magnetization M =1 corresponds to all spins equal to +1 ('black' sites) and there is only one way of obtaining this configuration, whose energy is exp(-32J/kT). In the terminology of black-white edges, this corresponds to A(8,0) = 1, yielding $A_{16}(M =1) = (-32J)$.

However, several the configuration states can yield M =0 i.e q can vary from 8 to 32 for p=8 (number of +1 spins = number of -1 spins) as shown in Table 1. Thus

$$A_{16}(M=0) - kT \ln \begin{pmatrix} 4356 + 768e^{\frac{8J}{kT}} + 1600e^{\frac{4J}{kT}} + 8e^{\frac{16J}{kT}} + 2e^{\frac{-32J}{kT}} + 64e^{\frac{-20J}{kT}} \\ + 120e^{\frac{-16J}{kT}} + 576e^{\frac{-12J}{kT}} + 2112e^{\frac{-8J}{kT}} + 3264e^{\frac{-4J}{kT}} \end{pmatrix}$$

Equating the two free energies, the transcendental equation given below results viz.

$$\ln \begin{pmatrix} 4356 + 768e^{\frac{8J}{kT_c}} + 1600e^{\frac{4J}{kT_c}} + 8e^{\frac{16J}{kT_c}} + 2e^{\frac{-32J}{kT_c}} + 64e^{\frac{-20J}{kT_c}} \\ + 120e^{\frac{-16J}{kT_c}} + 576e^{\frac{-12J}{kT_c}} + 2112e^{\frac{-8J}{kT_c}} + 3264e^{\frac{-4J}{kT_c}} \end{pmatrix} - \frac{32J}{kT_c} = 0 \quad (8)$$

The solution of the above equation(8) using MATLAB leads to $J/kT_c$ as 0.3117. The above equation is not altered if the differences in energies from all -1 spins are considered. This value deviates significantly from the anticipated value of 0.4407 for $J/kT_c$. When N = 36, the analogous A(p,q) values leads to a more algebraic equation and $J/kT_c$ in this case follows as 0.3730. Interestingly, this value is nearer to the exact value of 0.4407 than that obtained from the Bethe approximation for infinite lattice ($J/kT_c$ = 0.3466). When viewed from this perspective, it follows that an exact analysis pertaining to N =36 yields a more accurate value for $T_c$ without following the magnetization route. Furthermore, the magnitude of the critical temperature is dictated by the weights pertaining to each black-white edge and theorems of algebraic combinatorics [35] may become valuable in decoding the influence of the lattice size on the magnitude of critical temperatures. The extension of this procedure to N = 64 is precluded by the tedious computations involved for A (p,q) values. The above procedure for estimating $T_c$ is more suited when the number of sites is even.

**Magnetization for non-zero magnetic field at the thermodynamic limit**

Although the spontaneous magnetization vanishes for finite lattices, the partition function expressions are useful to deduce the magnetization for H ≠ 0. The equations for the magnetization arising from the differentiation of ln (Q) for N =16 and N =36 have been deduced using *MATHEMATICA* for H ≠ 0. For N =64, the *numerical differentiation* of ln Q obtained from the matrix leads to the dependence of magnetization on nearest neighbour interaction energies and magnetic field. A plot of M vs 1/N and the extrapolation to 1/N → 0 yields the M value in the thermodynamic limit. The satisfactory nature of such a methodology has been earlier demonstrated [36] to deduce the spontaneous magnetization for infinite lattices when H =0 while the present analysis employs an analogous method for H ≠ 0 wherein the plot of M vs 1/N is made for different values of kT/J and H; a typical plot of M Vs 1/N for kT/J = 2.9497 is given in Figure 3.

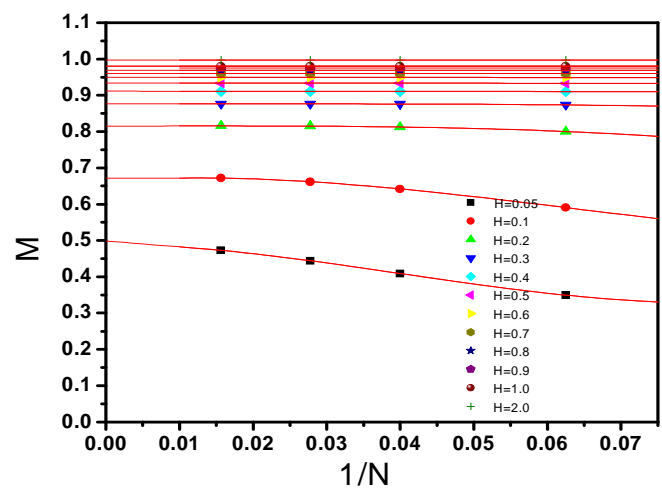

Figure 3 Extrapolation procedure for magnetization as N → ∞ for various values of H; the points denote the values obtained from differentiation of the partition function while the lines are obtained by the polynomial fitting.

The limit H → 0 was deduced subsequently for each kT/J ranging from 0.2269 to 7.4877 i.e T/$T_c$ spanning the values from 0.1 to 3.3. It is of interest to investigate whether any *empirical equation* exists for predicting the values of magnetization. By modifying the Bragg-Williams equation with two empirical constants, the following equation was found satisfactory for *fitting* the magnetization values:

$$M = \tanh\left[H + M\frac{T_c}{T} + \sinh\left(2.061H\left(\frac{T_c}{T}\right)^{1.895}\right)\right] \quad (9)$$

Figures 4a to 4c depict the magnetization obtained by this method for T < $T_c$; T = $T_c$; T > $T_c$; respectively. Figure 4d depicts the dependence of the magnetization on the magnetic field for various kT/J values. Figures 4a to 4d are consistent with the qualitative variation predicted by Huang [28] and Baxter [37]. In view of the availability for magnetization for various magnetic fields, it is imperative to deduce the critical exponent δ. For this purpose, the variation of ln M with ln H is analyzed when T/$T_c$ is ~ 1. As shown in Figure 6, the value of 1/δ is found to be 0.06282 ± 0.0044 in gross agreement[38] with the critical exponent of two-dimensional Ising models viz. i.e. 1/δ =0.0667 .

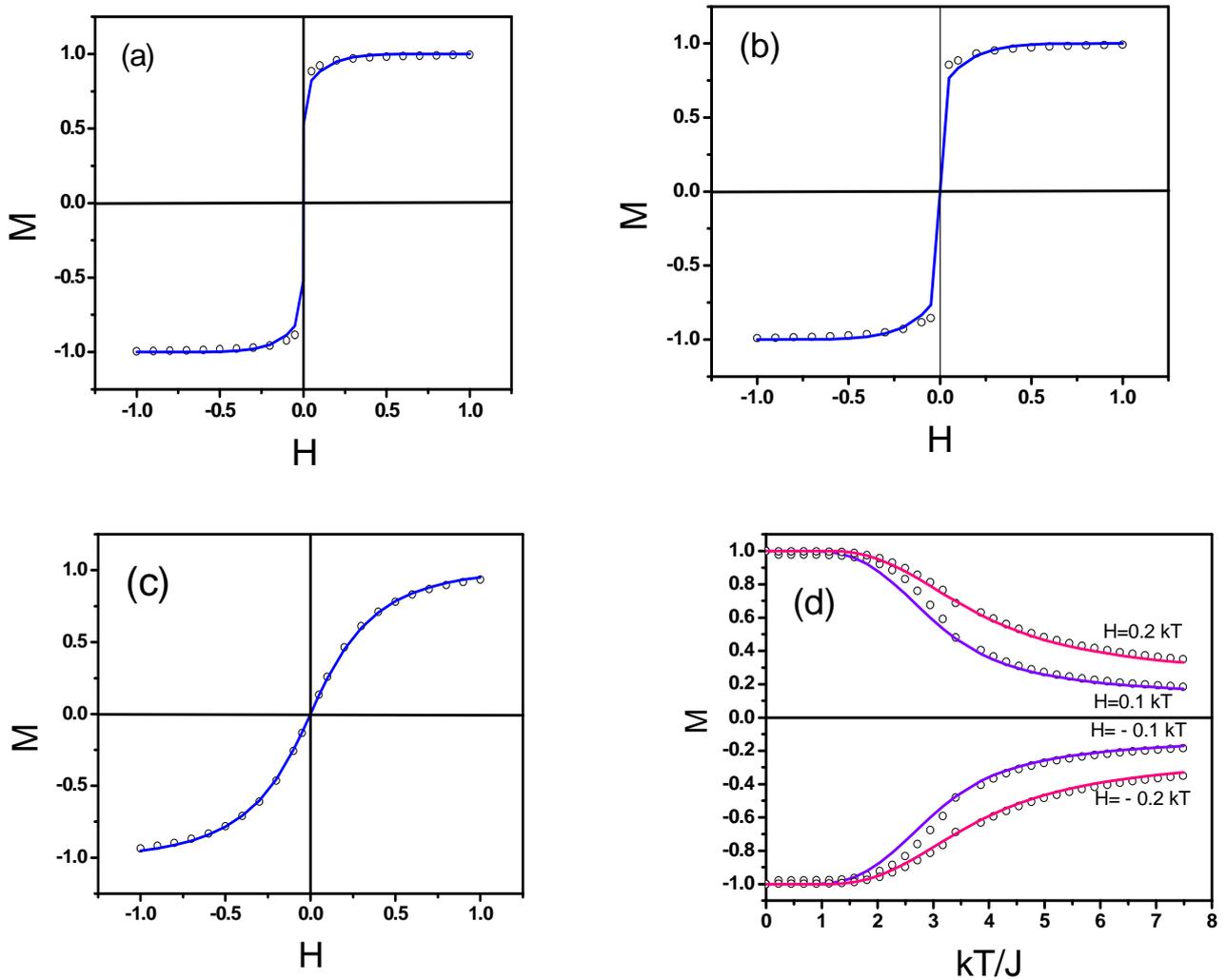

Figure 4 The variation of the spontaneous magnetization with the magnetic field for infinite lattice: (a) $T<T_c$, (b) $T=T_c$, (c) $T>T_c$ ; (d) Dependence of M on kT/J for various values of H for infinite lattice. The points denote the extrapolated values while the line denotes the prediction of equation 9.

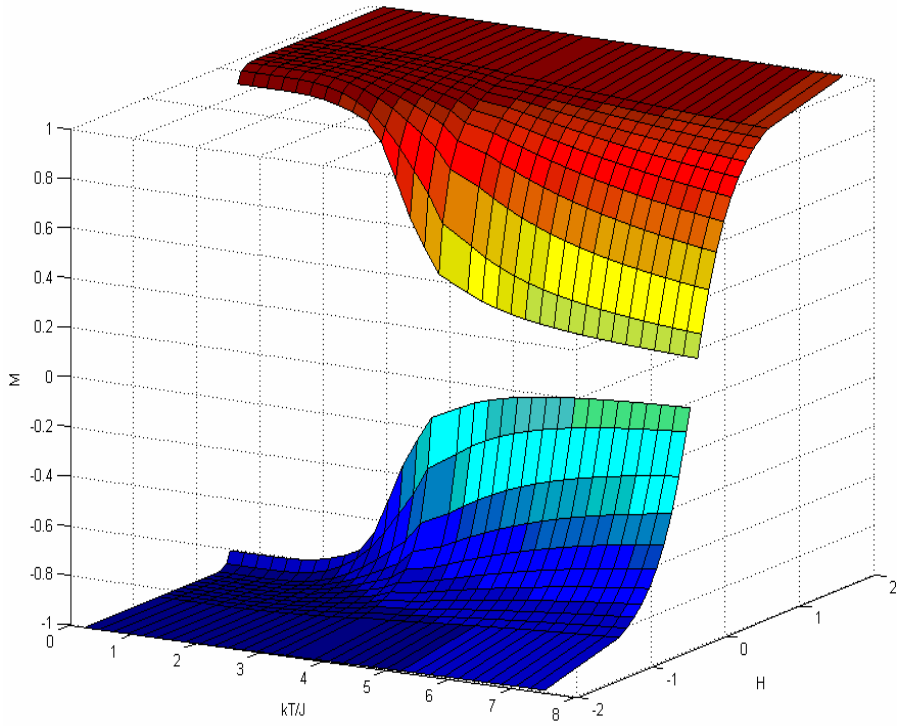

Figure 5 -The variation of magnetization with magnetic field and nearest neighbour interaction energy arising from the values pertaining to infinite lattice.

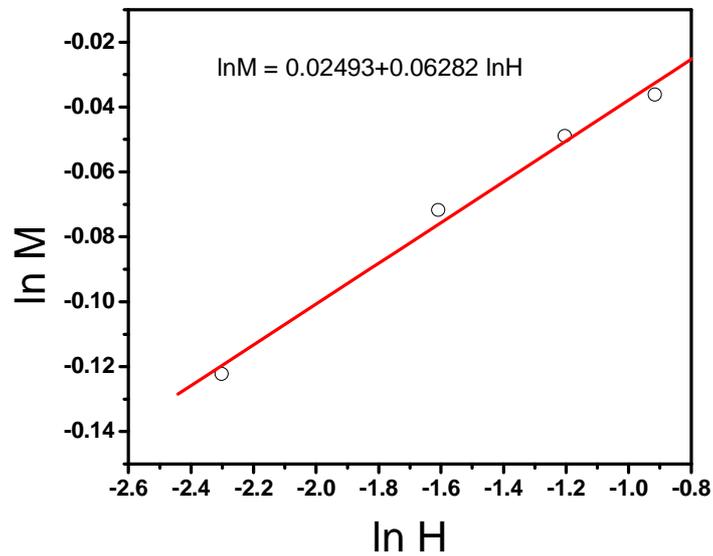

ln M = 0.02493+0.06282 lnH

Figure 6 The variation of ln M with ln H at the critical temperature. The points denote the computed values of M while the line is obtained from the linear regression analysis.

**Magnetic Susceptibility**

The analysis of magnetic susceptibility of Ising models is very intricate in view of the sensitive nature of the series coefficients as was demonstrated by Fisher [39] and Sykes [40]. Consequently, our present analysis focuses only on the susceptibility values for *non-zero* magnetic field for N =16, N =36 and N =64.

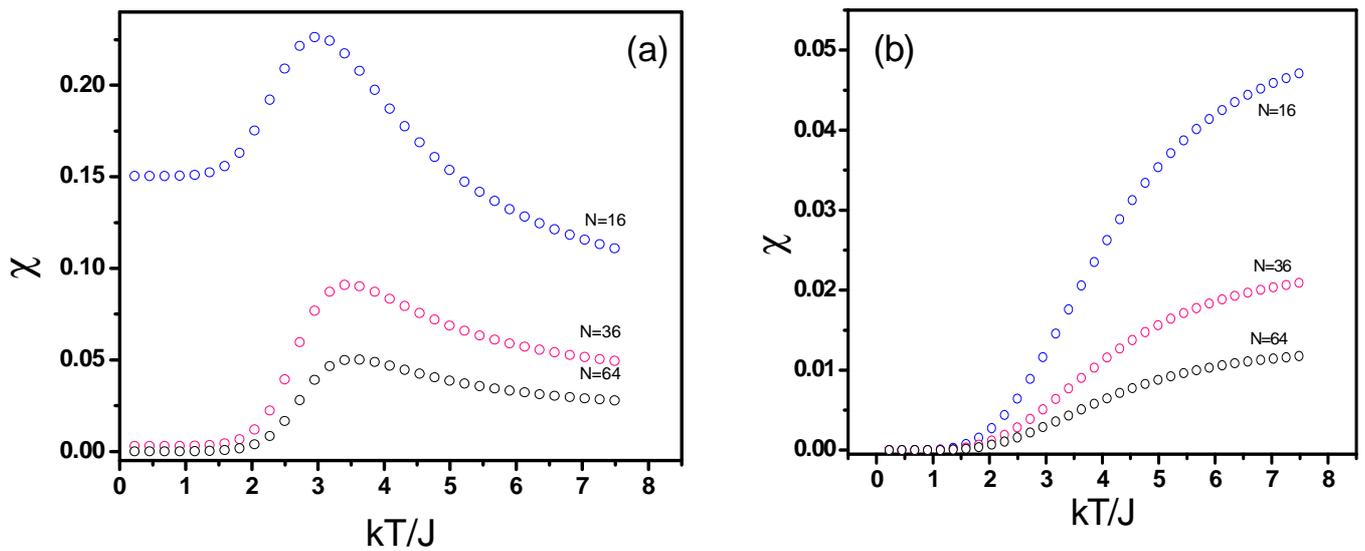

Figure 7 The dependence of the susceptibility ($\chi$) on (kT/J) for N=16, 36 and 64 at different magnetic fields :(a) H=0.1 and (b) H=0.5

Since the magnetization pertaining to the infinite lattice itself has been deduced from the extrapolation procedure, differentiation of the same leads to the variation of the susceptibility predicted in Figure 8. In view of the sensitivity of the numerical differentiation of the extrapolated data, these estimates may involve an error of *ca.* 5%. It is of interest to note from Figure 8 that the maximum in susceptibility arises only when H $\rightarrow$ 0. As mentioned earlier, the magnetization equation itself is not employed to deduce the susceptibility on account of the inaccuracies associated with the mathematical operations of empirical equations.

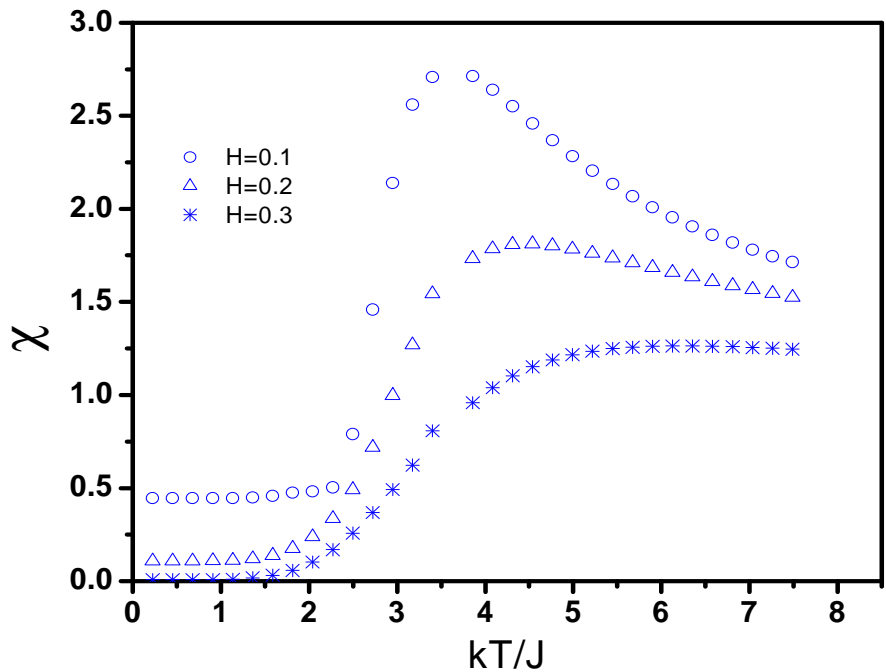

Figure 8 The dependence of the susceptibility (χ) on nearest neighbor interaction energy (kT/J) for various magnetic fields employing the magnetization estimates of the infinite lattice.

**Specific Heat**

The specific heat of two-dimensional Ising models constitutes a sensitive test of validating any computational procedure [11]. The specific heat is related to the internal energy as

$$U = kT^2 \frac{\partial \ln Q}{\partial T}$$

$$C_v = \frac{\partial U}{\partial T}$$

Employing the ln Q values for N = 16, N= 36 and N =64 at H =0, the predicted variation of $C_v$ with kT/J is depicted in Figure 9(a) for H =0.While the $C_v$ variation for 16 and 36 sites were obtained by analytical differentiation of the partition function expression, the Cv variation pertaining to 8×8 lattice was obtained by numerical differentiation of the partition function for 64 sites. An empirical equation pertaining to $C_v$ for H ≠ 0 is given below:

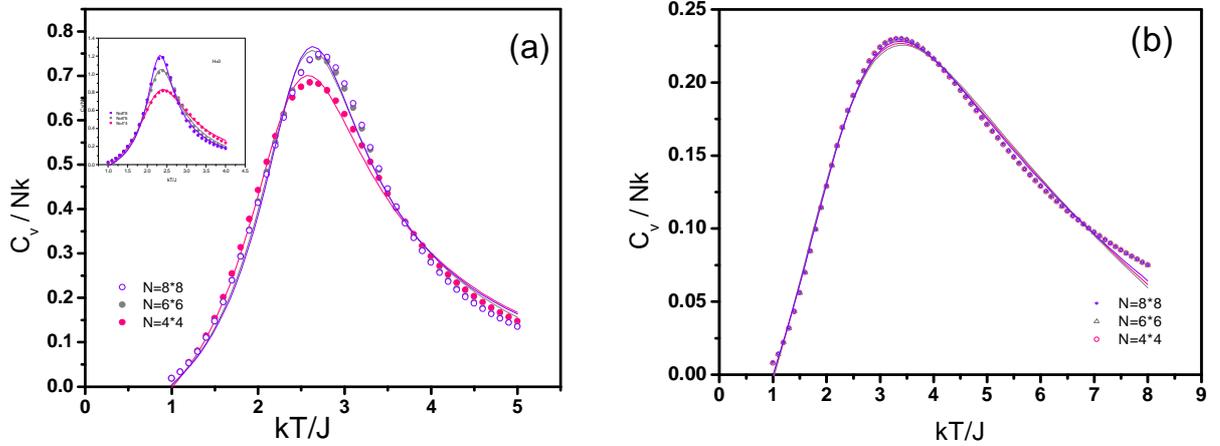

Figure 9 The dependence of the specific heat on (kT/J) for N=16, 36 and 64 at different magnetic fields :(a) H=0.1 and (b) H=0.5.The points denote the values arising from the partition function while the line denotes the prediction of equation (10).The inset depicts the specific heat at H =0.

$$\frac{C_v}{Nk} = a\frac{T}{T_c}\ln\left[\left(\frac{T}{T_c}-1+\left(\frac{b}{n}+a_1 H^2\frac{T}{T_c}\right)\right)^2 + \left(\frac{c^2}{n^2}+b_1 H^2\frac{T}{T_c}\right)+c_1 H^2\right]^{-\frac{1}{2}} + d\left(\frac{T}{T_c}+1+d_1 H^2\right) \quad (10)$$

Equation (10) when H =0 (i) leads to the singularity at the critical temperature and (ii) is similar to the expression reported by Ferdinand and Fisher [41]. Figure 9(b) depicts the variation of $C_v$ with nearest neighbour interaction energy for a few typical values of the magnetic field.

Table 4 - The numerical values of the constants in equation (10) at H =0

| N  | a              | b               | c              | d               |
|----|----------------|-----------------|----------------|-----------------|
| 16 | 0.2958±0.0171  | 0.0059±0.0190   | 4.1368±0.0110  | -0.1693±0.0147  |
| 36 | 0.2414±0.0146  | -0.0449±0.0226  | 6.1000±0.0110  | -0.1502±0.0148  |
| 64 | 0.20144±0.0114 | -0.0673±0.0244  | 8.0797±0.0101  | -0.1236±0.0130  |

Table 5 - The numerical values of the constants in equation (10) at H =0.1

| N  | a1             | b1              | c1             | d1              |
|----|----------------|-----------------|----------------|-----------------|
| 16 | 8.3141±0.0025  | -32.9655±0.7652 | 36.7009±0.8376 | -28.0556±2.0525 |
| 36 | 0.7151±0.0373  | -20.4327±0.7904 | 25.4767±0.8922 | -37.0773±2.9387 |
| 64 | -9.4666±0.3572 | 0.9658±0.0126   | 2.3181±0.1483  | -26.8382±4.7988 |

Table 6 - The numerical values of the constants in equation (10) at H =0.5

| N  | a1             | b1             | c1             | d1             |
|----|----------------|----------------|----------------|----------------|
| 16 | -0.3481±0.1013 | 0.5975±0.2923  | 1.5739±0.4154  | -1.8417±0.1530 |
| 36 | -1.1972±0.0966 | 1.8462±0.1670  | -0.7412±0.3101 | -1.3674±0.2239 |
| 64 | -1.176±0.0637  | 1.5530±0.1156  | -0.6219±0.2006 | -1.3338±0.1586 |

It is intriguing to note that the vanishing of the spontaneous magnetization, maximum in susceptibility as well as its temperature derivative all occur at the same point viz Tc. This arises on account of the symmetry underlying the magnetization (order parameter) equation. Since the order parameter is an odd function of the magnetic field, the susceptibility viz (dM/dH) is an even function of H. The partial derivative of magnetization with respect to temperature is also an odd function while the derivative with respect to H is an even function. Thus, it follows that the susceptibility and its temperature derivative will have the extrema at the same temperature viz Tc, where the spontaneous magnetization becomes zero. This insight has an interesting connotation in the statistical mechanical theory of the electrical double layer wherein the vanishing of the dipole potential and the extrema of the differential capacitance and its temperature coefficient occur at the same surface charge density [42].While the sizes of the lattice chosen here are quite small in comparison with the Monte Carlo simulations [36] and other numerical procedures [43], the present strategy is entirely new since all the thermodynamic quantities are obtained from the corresponding partition functions-a central quantity in statistical mechanics. Furthermore, the partition functions reported herein for $N = 16$, $N = 25$, $N = 36$ and $N = 64$ are exact, these values being deduced for zero and non-zero magnetic fields. While the recurrence relation needs further refinement in order to yield accurate thermodynamic parameters, the algebraic expressions for the canonical partition functions are reported for the first time. It appears therefore that the Holy Grail is not entirely un-surmountable.

In Summary, the partition functions for two-dimensional Ising models for small square lattices in the presence of the magnetic field are computed by different methods and the magnetization, specific heat and susceptibilities are deduced therefrom. The extrapolation to infinite lattice sizes for magnetization in the presence of the magnetic field is suggested. The critical temperatures are deduced for $N = 16$ and $N = 36$ using graph theoretical procedures while the magnetization critical exponent $\delta$ is computed from the magnetization values.

**REFERENCES**

1. Chamberlin, R.V. Mean-field model for the critical behaviour of ferromagnets. *Nature* **408,** 337-339 (2000).

2. Peryard, M. Melting the double helix, *Nature Phys* **2**, 13-14 (2006).

3. Onsager, L. Crystal Statistics. I. A Two-Dimensional Model with an Order-Disorder Transition. *Phys.Rev* **65**,117-149 (1944).

4. Lee, T.D. & Yang, C.N. Statistical theory of equations and phase transitions. II. Lattice gas and Ising Model. *Phys. Rev* **87,** 410-419 (1952).

5. Bragg, W.L. & Williams, E.J. The effect of thermal agitation on atomic arrangement in alloys. *Proc. Roy. Soc. A* **145**, 699-730 (1934).

6. Bethe, H.A. Statistical theory of superlattices. *Proc. Roy. Soc. A* **150**, 552-575 (1935).

7. Domb, C. In Domb, C. & Green, M. S., *Phase Transitions and Critical Phenomena*. (Academic Press, 1974).

8. Wilson, K.G. Renormalization group and critical phenomena. I. Renormalization group and the Kadanoff scaling picture. *Phys.Rev.B* **4**, 3174-3205 (1971).

9. Kadanoff, L.P. *et al.* Static Phenomena Near Critical Points: Theory and Experiment. *Rev.Mod.Phys* **39,** 395-437 (1967).

10. Hill, T. L. *Statistical Mechanics – Principles and Selected Applications* (McGraw-Hill Book Company, 1956).

11. McCoy, B.M. & Wu, T.T. *Two-Dimensional Ising Model* (Harvard University Press, 1973).

12. Pushpalatha, K. & Sangaranarayanan.M.V. Two-dimensional condensation of organic adsorbates at the mercury/aqueous solution interface: A global analysis of critical parameters using Ising model formalism. *J.Electroanal.Chem* **425,** 39-48 (1997).

13. Saradha, R. & Sangaranarayanan, M. V. Theory of electrified interfaces: A combined analysis using density functional approach and Bethe Approximation. *J.Phys.Chem B* **102,** 5468-5474 (1998).

14. Aoki, K. Theory of phase separation of binary self-assembled films. *J.Electroanal.Chem* **513,** 1-7 (2001).

15. Lenz, P., Zagrovic, B., Shapiro, J. & Pande, V.S. Folding probabilities: A novel approach to folding transitions and the two-dimensional Ising-model. *J.Chem.Phys* **120,** 6769-6778 (2004).

16. Kole, P. R., de Vries, R. J., Poelsema, B. & Zandvliet, H. J. W. Free Energies of steps on (1 1 1) fcc surfaces. *Solid State Comm* **136,** 356-359 (2005).

17. Thompson, C. J., *The Annals of Probability* **14**, 1129-1138 (1986).

18. Plischke, M. & Bergersen, B. *Equilibrium Statistical Physics* (Prentice-Hall, 1989).

19. Chandler, D., *Introduction to Statistical Mechanics* (Oxford University Press, 1987).

20. Newman, M.E.J & Barkema, G.T., *Monte Carlo Methods in Statistical Physics* (Oxford University Press, 1998).

21. Kasteleyn, P.W., Graph Theory and Crystal Physics. (Academic Press, 1967).

22. Chang, S-C & Shrock R., Weighted Graph Colorings. *J.Stat.Phys* **138,** 496-542 (2009).

23. Finch, S.R., *Mathematical Constants* (Cambridge University Press, 2003).

24. Dokholyan, N.V., Li, L., Ding, F., Shakhnovich, E.I., Topological determinants of protein folding. *Proc. Natl. Acad. Sci* **99,** 8637-8641 (2002).

25. Bodroza-Pantic, O., The Gutman formulas for algebraic structure count. *J.Math.Chem* **35,** 139-146 (2004).

26. Maayan, A., Insights into the Organization of Biochemical Regulatory Networks Using Graph Theory Analyses. *J.Biol.Chem* **284,** 5451-5455 (2009).

27. Temperley, H.N.V., & Lieb, E.M., Relations between the 'percolation' and coloring problem and other graph theoretical problems associated with regular lattices: some exact results for the 'percolation' problem. *Proc.Roy.Soc.Lond.A* **322**, 251-280 (1971).

28. Huang. K., Statistical Mechanics. (John Wiley & Sons, 1963).

29. Kramers, H.A., & Wannier G.H., Statistics of the Two-Dimensional Ferromagnet. Part I. *Phys.Rev* **60**, 252-262 (1941).

30. Nandhini, G & Sangaranarayanan, M.V., Partition Function of Nearest neighbour Ising models: Some new insights. *J.Chem.Sci* **121**, 595-599 (2009).

31. Noor, F & Morgera, S.D., Construction of a Hermitian Toeplitz Matrix from an arbitrary set of Eigenvalues. *IEEE Trans. Signal processing* **40**, 2093-2094 (1992).

32. Bellman, R., Introduction to Matrix Analysis. (AM publishing House, 1997).

33. Haggkvist, R., Rosengren, A., Andren, D., Kundrotas, P., Lundow, P.H., Markstrom, K., Computation of the Ising Partition function for 2-dimensional square grids. *Phys.Rev.E* **69**, 046104 (2004).

34. Newell, G.F. & Montroll, E.W., On the Theory of the Ising Model of Ferromagnetism. *Rev.Mod.Phys* **25**, 353-389 (1953).

35. Gimbel, J., & Kennedy, J.W., Quo Vadis Graph Theory?. (North-Holland, 1993).

36. Landau.D.P., Finite-Size behaviour of the Ising square lattice. *Phys.Rev.B* **13**, 2297-3011 (1976).

37. Baxter, R.J., Exactly Solved Models in Statistical Mechanics. (Academic Press Limited, 1982).

38. Salinas, S.R.A., Introduction to Statistical Physics. (Springer, 2004).

39. Fisher, M.E., The Susceptibility of Plane Ising Model. *Physica* **25,** 521-524 (1959).

40. Domb, C. & Sykes, M.F., On the Susceptibility of a Ferromagnetic above the Curie Point. *Proc. Roy. Soc. A* **240**, 214-228 (1957).

41. Ferdinand, A.E., & Fisher, M.E., Bounded and Inhomogenous Ising Models. I. Specific Heat anomaly of a finite lattice. *Phys.Rev* **185**, 832-846 (1969).

42. Rangarajan, S.K., Specialist Periodical Reports, Electrochemistry, **7**, The Chemical Society, London. (1980).

43. Potamianos, G.G., & Goutsias, J.K., Partition Function Estimation of Gibbs Random Field Images Using Monte Carlo Simulations. *IEEE Trans. Inform. Theory* **39**, 1322-1332 (1993).

Appendix-Typical values of A(p,q) for N=36 leading to the partition function of the two-dimensional Ising model

A(0,0)   1

A(1,4)   36

A(2,6)   72

A(2,8)   558

A(3,8)   216

A(3,10)   2016

A(3,12)   4908

A(4,8)   36

A(4,10)   648

A(4,12)   7308

A(4,14)   23688

A(4,16)   27225

A(5,10)   288

A(5,12)   2844

A(5,14)   24480

A(5,16)   95544

A(5,18)   153504

A(5,20)   100332

A(6,10)   72

A(6,12)   1452

A(6,14)   13968

A(6,16)   86544

A(6,18)   336024

A(6,20)   648828

A(6,22)   608112

A(6,24)   252792

A(7,12)   792

A(7,14)   7632

A(7,16)   62136

A(7,18)   320976

A(7,20)   1118268

A(7,22)   2282256

A(7,24)   2564496

A(7,26)   1548144

A(7,28)   442980

A(8,12)   216

A(8,14)   4968

A(8,16)   41634

A(8,18)   258408

A(8,20)   1169172

A(8,22)   3595176

A(8,24)   7164882

A(8,26)   8636976

A(8,28)   6240528

A(8,30)   2601864

A(8,32)   546516

A(9,12)   36

A(9,14)   2736

A(9,16)   28368

A(9,18)   203232

A(9,20)   1040544

A(9,22)   4035024

A(9,24)   11037624

A(9,26)   20542608

A(9,28)   24871536

A(9,30)   19324416

A(9,32)   9645120

A(9,34)   2931120

A(9,36)   480916

A(10,14)   1224

A(10,16)   19764

A(10,18)   155160

A(10,20)   898164

A(10,22)   3950064

A(10,24)   13086414

A(10,26)   31673808

A(10,28)   53905896

| | |
|---|---|
| A(10,30) | 62584632 |
| A(10,32) | 49293576 |
| A(10,34) | 26428248 |
| A(10,36) | 9647748 |
| A(10,38) | 2233584 |
| A(10,40) | 308574 |
| A(11,14) | 432 |
| A(11,16) | 13464 |
| A(11,18) | 118944 |
| A(11,20) | 761616 |
| A(11,22) | 3683376 |
| A(11,24) | 13794192 |
| A(11,26) | 39356208 |
| A(11,28) | 83537208 |
| A(11,30) | 127874304 |
| A(11,32) | 138281976 |
| A(11,34) | 105570288 |
| A(11,36) | 57644748 |
| A(11,38) | 22527648 |
| A(11,40) | 6331896 |
| A(11,42) | 1158048 |
| A(11,44) | 150948 |
| A(12,12) | 12 |

A(12,14)   72

A(12,16)   8820

A(12,18)   93384

A(12,20)   642528

A(12,22)   3363048

A(12,24)   13788678

A(12,26)   43856208

A(12,28)   107681508

A(12,30)   198847464

A(12,32)   269648586

A(12,34)   265556736

A(12,36)   190654740

A(12,38)   101625120

A(12,40)   40540428

A(12,42)   12133104

A(12,44)   2765664

A(12,46)   410832

A(12,48)   60768

A(13,14)   144

A(13,16)   5184

A(13,18)   74592

A(13,20)   550836

A(13,22)   3051504

| | |
|---|---|
| A(13,24) | 13410360 |
| A(13,26) | 46237392 |
| A(13,28) | 125152164 |
| A(13,30) | 262657728 |
| A(13,32) | 417699720 |
| A(13,34) | 494848800 |
| A(13,36) | 435314160 |
| A(13,38) | 287356320 |
| A(13,40) | 145123416 |
| A(13,42) | 56807136 |
| A(13,44) | 17364096 |
| A(13,46) | 4192704 |
| A(13,48) | 820224 |
| A(13,50) | 101520 |
| A(13,52) | 21600 |
| A(14,14) | 144 |
| A(14,16) | 3456 |
| A(14,18) | 60840 |
| A(14,20) | 475092 |
| A(14,22) | 2803824 |
| A(14,24) | 12893472 |
| A(14,26) | 47208816 |
| A(14,28) | 137160000 |
| A(14,30) | 313764192 |
| A(14,32) | 557542728 |
| A(14,34) | 753788160 |
| A(14,36) | 769393656 |
| A(14,38) | 594585792 |
| A(14,40) | 354531024 |
| A(14,42) | 165822984 |
| A(14,44) | 61849620 |
| A(14,46) | 18671616 |
| A(14,48) | 4592088 |
| A(14,50) | 951552 |
| A(14,52) | 173304 |
| A(14,54) | 18144 |
| A(14,56) | 6696 |
| A(15,14) | 144 |
| A(15,16) | 2808 |
| A(15,18) | 50160 |
| A(15,20) | 423540 |
| A(15,22) | 2599344 |
| A(15,24) | 12408816 |
| A(15,26) | 47457792 |
| A(15,28) | 144748116 |
| A(15,30) | 351857376 |
| A(15,32) | 673475688 |
| A(15,34) | 997994592 |
| A(15,36) | 1131880752 |
| A(15,38) | 979974720 |
| A(15,40) | 657496440 |
| A(15,42) | 348192192 |
| A(15,44) | 148241772 |
| A(15,46) | 51576336 |
| A(15,48) | 14895720 |
| A(15,50) | 3683376 |
| A(15,52) | 757296 |
| A(15,54) | 153456 |
| A(15,56) | 28152 |
| A(15,58) | 2304 |
| A(15,60) | 1668 |
| A(16,14) | 144 |
| A(16,16) | 2700 |
| A(16,18) | 42120 |
| A(16,20) | 395568 |
| A(16,22) | 2425752 |
| A(16,24) | 12095820 |
| A(16,26) | 47250864 |
| A(16,28) | 149403996 |

A(16,30)   377578800

A(16,32)   758452707

A(16,34)   1193279472

A(16,36)   1450906884

A(16,38)   1356521904

A(16,40)   985754664

A(16,42)   566862840

A(16,44)   263218752

A(16,46)   100316088

A(16,48)   32060817

A(16,50)   8660952

A(16,52)   2091492

A(16,54)   443016

A(16,56)   82404

A(16,58)   20448

A(16,60)   3456

A(16,62)   144

A(16,64)   306

A(17,14)   144

A(17,16)   2664

A(17,18)   38592

A(17,20)   373176

A(17,22)   2347344

A(17,24)   11825928

A(17,26)   47161728

A(17,28)   151680312

A(17,30)   392376672

A(17,32)   810164736

A(17,34)   1317182688

A(17,36)   1668054600

A(17,38)   1628668800

A(17,40)   1240315488

A(17,42)   747933120

A(17,44)   364675500

A(17,46)   146534832

A(17,48)   49325472

A(17,50)   14213952

A(17,52)   3589416

A(17,54)   806544

A(17,56)   182016

A(17,58)   33408

A(17,60)   6912

A(17,62)   2304

A(17,64)   216

A(17,68)   36

A(18,12)   12

A(18,16)   3456

A(18,18)   34704

A(18,20)   373752

A(18,22)   2299680

A(18,24)   11788038

A(18,26)   47010240

A(18,28)   152525196

A(18,30)   397078512

A(18,32)   827402400

A(18,34)   1359828000

A(18,36)   1744323016

A(18,38)   1728000432

A(18,40)   1336101246

A(18,42)   818194560

A(18,44)   405640980

A(18,46)   165466512

A(18,48)   56924808

A(18,50)   16566624

A(18,52)   4302864

A(18,54)   1004256

A(18,56)   205542

A(18,58)   47520

A(18,60)   11076

A(18,62)   1152

A(18,64)   576

A(18,66)   144

A(18,72)   2

A(19,14)   144

A(19,16)   2664

A(19,18)   38592

A(19,20)   373176

A(19,22)   2347344

A(19,24)   11825928

A(19,26)   47161728

A(19,28)   151680312

A(19,30)   392376672

A(19,32)   810164736

A(19,34)   1317182688

A(19,36)   1668054600

A(19,38)   1628668800

A(19,40)   1240315488

A(19,42)   747933120

A(19,44)   364675500

A(19,46)   146534832

A(19,48)   49325472

A(19,50)   14213952

A(19,52)  3589416

A(19,54)  806544

A(19,56)  182016

A(19,58)  33408

A(19,60)  6912

A(19,62)  2304

A(19,64)  216

A(19,68)  36

A(20,14)  144

A(20,16)  2700

A(20,18)  42120

A(20,20)  395568

A(20,22)  2425752

A(20,24)  12095820

A(20,26)  47250864

A(20,28)  149403996

A(20,30)  377578800

A(20,32)  758452707

A(20,34)  1193279472

A(20,36)  1450906884

A(20,38)  1356521904

A(20,40)  985754664

A(20,42)  566862840

A(20,44)   263218752

A(20,46)   100316088

A(20,48)   32060817

A(20,50)   8660952

A(20,52)   2091492

A(20,54)   443016

A(20,56)   82404

A(20,58)   20448

A(20,60)   3456

A(20,62)   144

A(20,64)   306

A(21,14)   144

A(21,16)   2808

A(21,18)   50160

A(21,20)   423540

A(21,22)   2599344

A(21,24)   12408816

A(21,26)   47457792

A(21,28)   144748116

A(21,30)   351857376

A(21,32)   673475688

A(21,34)   997994592

A(21,36)   1131880752

A(21,38)   979974720

A(21,40)   657496440

A(21,42)   348192192

A(21,44)   148241772

A(21,46)   51576336

A(21,48)   14895720

A(21,50)   3683376

A(21,52)   757296

A(21,54)   153456

A(21,56)   28152

A(21,58)   2304

A(21,60)   1668

A(22,14)   144

A(22,16)   3456

A(22,18)   60840

A(22,20)   475092

A(22,22)   2803824

A(22,24)   12893472

A(22,26)   47208816

A(22,28)   137160000

A(22,30)   313764192

A(22,32)   557542728

A(22,34)   753788160

A(22,36)   769393656

A(22,38)   594585792

A(22,40)   354531024

A(22,42)   165822984

A(22,44)   61849620

A(22,46)   18671616

A(22,48)   4592088

A(22,50)   951552

A(22,52)   173304

A(22,54)   18144

A(22,56)   6696

A(23,14)   144

A(23,16)   5184

A(23,18)   74592

A(23,20)   550836

A(23,22)   3051504

A(23,24)   13410360

A(23,26)   46237392

A(23,28)   125152164

A(23,30)   262657728

A(23,32)   417699720

A(23,34)   494848800

A(23,36)   435314160

A(23,38)   287356320

A(23,40)   145123416

A(23,42)   56807136

A(23,44)   17364096

A(23,46)   4192704

A(23,48)   820224

A(23,50)   101520

A(23,52)   21600

A(24,12)   12

A(24,14)   72

A(24,16)   8820

A(24,18)   93384

A(24,20)   642528

A(24,22)   3363048

A(24,24)   13788678

A(24,26)   43856208

A(24,28)   107681508

A(24,30)   198847464

A(24,32)   269648586

A(24,34)   265556736

A(24,36)   190654740

A(24,38)   101625120

A(24,40)   40540428

A(24,42)   12133104

A(24,44)   2765664

A(24,46)   410832

A(24,48)   60768

A(25,14)   432

A(25,16)   13464

A(25,18)   118944

A(25,20)   761616

A(25,22)   3683376

A(25,24)   13794192

A(25,26)   39356208

A(25,28)   83537208

A(25,30)   127874304

A(25,32)   138281976

A(25,34)   105570288

A(25,36)   57644748

A(25,38)   22527648

A(25,40)   6331896

A(25,42)   1158048

A(25,44)   150948

A(26,14)   1224

A(26,16)   19764

A(26,18)   155160

A(26,20)   898164

A(26,22)   3950064

A(26,24)   13086414

A(26,26)   31673808

A(26,28)   53905896

A(26,30)   62584632

A(26,32)   49293576

A(26,34)   26428248

A(26,36)   9647748

A(26,38)   2233584

A(26,40)   308574

A(27,12)   36

A(27,14)   2736

A(27,16)   28368

A(27,18)   203232

A(27,20)   1040544

A(27,22)   4035024

A(27,24)   11037624

A(27,26)   20542608

A(27,28)   24871536

A(27,30)   19324416

A(27,32)   9645120

A(27,34)   2931120

| | |
|---|---|
| A(27,36) | 480916 |
| A(28,12) | 216 |
| A(28,14) | 4968 |
| A(28,16) | 41634 |
| A(28,18) | 258408 |
| A(28,20) | 1169172 |
| A(28,22) | 3595176 |
| A(28,24) | 7164882 |
| A(28,26) | 8636976 |
| A(28,28) | 6240528 |
| A(28,30) | 2601864 |
| A(28,32) | 546516 |
| A(29,12) | 792 |
| A(29,14) | 7632 |
| A(29,16) | 62136 |
| A(29,18) | 320976 |
| A(29,20) | 1118268 |
| A(29,22) | 2282256 |
| A(29,24) | 2564496 |
| A(29,26) | 1548144 |
| A(29,28) | 442980 |
| A(30,10) | 72 |
| A(30,12) | 1452 |

A(30,14)   13968

A(30,16)   86544

A(30,18)   336024

A(30,20)   648828

A(30,22)   608112

A(30,24)   252792

A(31,10)   288

A(31,12)   2844

A(31,14)   24480

A(31,16)   95544

A(31,18)   153504

A(31,20)   100332

A(32,8)    36

A(32,10)   648

A(32,12)   7308

A(32,14)   23688

A(32,16)   27225

A(33,8)    216

A(33,10)   2016

A(33,12)   4908

A(34,6)    72

A(34,8)    558
A(35,4)    36
A(36,0)    1